\documentclass[aps,prd,tightenlines,nofootinbib,amsfonts,amssymb,amsmath,11pt]{revtex4}
\usepackage{graphicx} 
\usepackage{color}

\begin{document}
\newcommand{\etal}{{\it et al.}} 
\newcommand{\bx}{{\bf x}}
\newcommand{\bn}{{\bf n}} 
\newcommand{\bk}{{\bf k}}
\newcommand{\dd}{{\rm d}}
\newcommand{\dslash}{D\!\!\!\!/}
\def\ga{\mathrel{\raise.3ex\hbox{$>$\kern-.75em\lower1ex\hbox{$\sim$}}}}
\def\la{\mathrel{\raise.3ex\hbox{$<$\kern-.75em\lower1ex\hbox{$\sim$}}}}
\def\beq{\begin{equation}}
\def\eeq{\end{equation}}
\def\be{\begin{equation}} 
\def\ee{\end{equation}}
\def\bea{\begin{eqnarray}}
\def\eea{\end{eqnarray}}

\def\C{{\cal C}}
\def\cleeu{{\tilde \C}_l^{EE}}
\def\clteu{{\tilde \C}_l^{TE}}
\def\clttu{{\tilde \C}_l^{TT}}
\def\a{\alpha}
\def\im{\eta^{-1}}

\def\tr{{\rm tr}}

\vskip-2cm
\title{A scenario for inflationary magnetogenesis without strong coupling problem}

\author{Gianmassimo Tasinato}

\affiliation{Department of Physics, Swansea University, Swansea, SA2 8PP, UK}
\affiliation{Institute of Cosmology and Gravitation, University of Portsmouth, Portsmouth, 
PO1 3FX, UK
}

\begin{abstract}
Cosmological magnetic fields pervade the entire universe, from small to
large scales.  Since they apparently extend into the intergalactic
medium, it is tantalizing to believe that they have a primordial origin,
possibly being produced during inflation. However, finding consistent
scenarios for inflationary magnetogenesis is a challenging theoretical
problem.  The requirements to avoid an excessive production of
electromagnetic energy, and to avoid entering a strong coupling regime
characterized by large values for the electromagnetic coupling constant,
typically allow one to generate only a tiny amplitude of magnetic field
during inflation.  We propose a scenario for building gauge-invariant
models of inflationary magnetogenesis potentially free from these
issues.  The idea is to derivatively couple a dynamical scalar, not
necessarily the inflaton, to fermionic and electromagnetic fields during
the inflationary era.  Such couplings give additional freedom to control
the time-dependence of the electromagnetic coupling constant during
inflation.  This fact allows us to find conditions to avoid the strong 
coupling problems that affect many of the existing models of
magnetogenesis. We do not need to rely on a particular inflationary
set-up for developing our scenario, that might be applied to different
realizations of inflation.  On the other hand, specific requirements
have to be imposed on the dynamics of the scalar derivatively coupled to
fermions and electromagnetism, that we are able to satisfy in an
explicit realization of our proposal.

\end{abstract}

\maketitle
\section{Introduction}

Cosmological magnetic fields seem to pervade the entire universe, from large to small scales.  
The existence of magnetic fields  at intergalactic distances has
  been inferred by the  lack of GeV  $\gamma-$rays detection
  from blazars, astrophysical objects  
    that are known to produce   photons with energies in the TeV range.   
    Interactions with the intergalactic medium  should
     convert at least      
 part of these  high energy TeV $\gamma-$rays   into
   lower energy secondary charged particles, which then should decay  into  GeV photons.
     The latter are however not detected, and 
    the simplest explanation for   this fact 
        is
    the presence of an
     intergalactic magnetic field,   that deflects the  secondary charged particles  \cite{Neronov:1900zz}.  
The required amplitude for such magnetic field is
found in  \cite{Tavecchio:2010mk} to 
 be at least $10^{-15}$ Gauss. 
 In addition,  magnetic fields are also measured within galaxies. They are thought to be amplified by dynamo effects starting from a seed
 magnetic field, that should have an amplitude of at least 
$10^{-20}$ Gauss 
  to render the dynamo  mechanism efficient.  See \cite{Durrer:2013pga} for a recent review.

\smallskip

Generating cosmological magnetic fields of such strengths is a challenging theoretical problem. Since they seem to pervade  the intergalactic medium,
 it is tantalizing to believe that they have a primordial origin, possibly  being produced during inflation.
  On the other hand, Maxwell theory of electromagnetism is conformally invariant. This implies that it does not lead by
  itself to the production of a coherent background of electromagnetic modes in a Friedmann-Robertson-Walker
  cosmology, since the latter can be expressed in conformally flat coordinates. To generate a sizeable magnetic field
  during inflation, the conformal invariance of Maxwell action has to be broken. 
   The
 first attempts in this direction have been push forward by Turner and Widrow in \cite{Turner:1987bw}, by coupling  Maxwell electromagnetism to space-time curvature.
   However, the most promising models in their set-up are plagued by ghosts \cite{Himmetoglu:2008zp,Himmetoglu:2008hx}, and are therefore inconsistent. Another 
    interesting scenario has been proposed by Ratra \cite{Ratra:1991bn}  by kinetically coupling    
   the inflaton scalar  field to
    electromagnetism,   in such a way to break  conformal invariance yet preserving
     gauge invariance. A consequence of  this approach  is that 
     the electromagnetic 
   coupling constant is time-dependent, changing rapidly  during 
      inflation. This scenario, although appealing in its simplicity, has nevertheless to face serious theoretical 
      problems. The requirements   to avoid an excessive production  of electromagnetic energy 
      (strong backraction problem) and to avoid very large values for   the
        electromagnetic coupling constant  (strong coupling problem)   allow   one to produce
         only    a tiny amplitude
         of magnetic field   during inflation  \cite{Demozzi:2009fu}.
         In fact, 
       despite the efforts of  many different groups to improve on 
  Ratra's scenario,  due to these and other problems there are very few  models
  of  primordial magnetogenesis that can be regarded as fully convincing from a theoretical perspective. 
  See 
   \cite{Durrer:2013pga,Widrow:2011hs,Subramanian:2009fu}
for    reviews that also  provide detailed
 surveys of  existing models.

\smallskip

In this work we propose  a   scenario    
   for building gauge-invariant   models of primordial magnetogenesis,   
 potentially free from  the aftermentioned backreaction and strong coupling issues.
  The   idea is to {\it derivatively couple} a dynamical scalar field to fermions and electromagnetism
   during inflation. 
    Derivative scalar-vector couplings that preserve
   gauge invariance, although explored in modified gravity scenarios,
     to the best of our knowledge have not been investigated in the context of primordial
    magnetogenesis. We show that their inclusion induces a time-dependent  sound speed for  electromagnetic
    and fermionic modes. Moreover, and most importantly, derivative couplings 
        give additional freedom 
      to control the time-dependence of the electromagnetic coupling
      constant during inflation:   this    quantity now  depends
        also on the {\it time derivative} of the scalar background
       profile.    This fact allows us to find conditions to avoid the serious strong coupling issue that, as
        pointed out in  \cite{Demozzi:2009fu}, 
        affects many of the  realizations
       of Ratra's proposal.

We do not need to rely on explicit inflationary  models for developing  our ideas, that 
might   
 be applied 
to different set-ups.  
  On the other hand,   specific  
 conditions   have to imposed on the dynamics of the scalar  derivatively coupled  to the electromagnetic and
 fermionic actions. 
 The simplest realization of our mechanism requires   a 
  homogenous profile for  the 
  scalar field with  a large time derivative in the early stages of the inflationary era.
   We will be able to discuss  an explicit example that   satisfies  our requirements in a consistent way. 
\smallskip

We start presenting the action for our system, that includes derivative couplings between a scalar  and the electromagnetic and fermion fields. We then explain  in general terms  
 how this framework allows us  to avoid the strong coupling problem of inflationary magnetogenesis, provided 
 that conditions are imposed on the dynamics of the scalar. After a section of conclusions,  various technical appendixes explore  novel phenomenological
 aspects of this scenario, and discuss a  concrete realization able to satisfy our requirements.

\section{The set-up } 
\label{sec-vesc}

We consider a gauge invariant theory 
 for electromagnetism during inflation, that includes 
  derivative couplings between a scalar   $\varphi$  and 
  electromagnetic and fermion fields during inflation. For definiteness, we 
  can think to the fermion field  as the electron. 
      The scalar $\varphi$  
        has a     time-dependent homogeneous profile during inflation,
   and   is  not    necessarily  the  
       inflaton.
    We will
  specify   the  requirements we impose on the dynamics of  $\varphi$ once   we develop  our scenario. 
  
  \smallskip

    The gauge-invariant, 
ghost free action that we examine   couples the scalar  $\varphi$  to the electromagnetic and fermion
 fields as 
 \be
 {\cal S}_{tot}
\,=\,{\cal  S}_{em}+ {\cal  S}_{\psi}\,,
 \ee
 with
\be\label{inact}
{\cal S}_{em}\,=\, \int d^4 x\,\sqrt{-g}
\left[
- \frac{1}{4}\,{\cal B}(\varphi)\,
	F_{\mu \nu} F^{\mu\nu}
	-{\cal C}(\varphi)
	\,
\left[ 
	\partial_\mu\varphi\,
	F^\mu_{\,\,\,\rho} F^{\rho\nu} \,  \partial_\nu\varphi
\right]\right]\,,
 \ee
and
\bea\label{galferm}
{\cal S}_{\psi}&=&\frac12\,\int d^4 x\,\sqrt{-g}\,\big\{
 i \, \bar{\psi}\,\gamma^\mu\,{\cal D}_\mu\,\psi
-i\,k_0 \,
\partial^\mu \varphi\,\partial_\nu \varphi
\,\left(\bar{\psi}\,
\gamma^\nu\,
{\cal D}_\mu
\,\psi\right)+h.c.
\big\}
\,. 
\eea
The  ingredients that make a difference with respect to standard scenarios as \cite{Ratra:1991bn} are 
 gauge-invariant 
{\it derivative
interactions} between the scalar and electromagnetic and fermion fields:
these are the dimension-8 operators proportional respectively to ${\cal C}(\varphi)$ and $k_0$ in eqs \eqref{inact}  
and \eqref{galferm}.   
 A recent work that analyzes in detail the  interesting phenomenological consequences 
 of derivative couplings between the scalar and gauge fields is \cite{Giovannini:2013rza}.
   (See also our Appendix \ref{sec-pheno}.)  
 %
   
   But the essential
 role for developing our arguments
   is played by  the new 
    coupling with matter. 
   That is, the 
    novel gauge invariant derivative operator proportional to $k_0$ in the fermionic action  \eqref{galferm} \footnote{We could  promote
   $k_0$ to a function of the scalar $\varphi$, without  
   qualitatively changing  our arguments.
    We will comment on this possibility towards the end of Section \ref{sec-strong}.}.
  Gauge-invariant derivative 
   couplings with this structure, although occasionally explored in modified gravity scenarios,
     to the best of our knowledge have not been investigated in the context of primordial
    magnetogenesis.
    
    We denote with  $F_{\mu\nu}\,=\,\partial_\mu A_\nu-\partial_\nu A_\mu$ the field strenght for a gauge field $A_\mu$.
      Being the electromagnetic action \eqref{inact} written in terms of $F_{\mu\nu}$,  it    
   is invariant under the abelian symmetry $A_\mu \to A_\mu+\partial_\mu \xi$ for arbitrary function $\xi$.  The 
 second operator proportional to ${\cal C}(\varphi)$ is of  dimension 8, but it is allowed 
  by the symmetries of the system, and as we shall see it  can become 
  important if the scalar acquires  a non-trivial
   time-dependent profile.

 The fermionic action \eqref{galferm} is also gauge invariant. We denote with  
 ${\cal D}_\mu=\nabla_\mu-i\,e\,A_\mu$  the covariant derivative, with $\nabla_\mu$ the space-time derivative in 
    curved space.
   Under a $U(1)$ transformation
 of the gauge field, $A_\mu\to A_\mu+\partial_\mu \xi$,  
 the fermionic field $\psi$ and its covariant derivative transform under the $U(1)$ gauge symmetry as
\bea
\psi&\to&e^{i \,e\,\xi}\,\psi
\,,
\\
\bar{\psi}&\to&e^{-i \,e\,\xi}\,\bar{\psi}
\,,
\\
{\cal D}_\mu \psi &\to&e^{i \,e\,\xi} \,{\cal D}_\mu \psi 
\,,
\eea
where $\xi(x)$ is an arbitrary function  controlling the gauge transformation.
 Hence $S_{el}$ is manifestly gauge invariant, including the derivative  dimension-8 operator proportional to $k_0$.  
  It is also not difficult to show that the complete action ${\cal S}_{tot}$
 is ghost free, since  $A_0$ remains a constraint and does not propagate.

Being  interested on inflationary cosmology, we   
 work with a conformally flat FRW metric, \be \label{cfmet}d s^2\,=\,a^2(\eta)\left[-d \eta^2 + d \vec{x}^2\right]\,.\ee
  For simplicity,
 we   focus our attention to the pure de Sitter case, where 
  the scale factor is $a(\eta)\,=\,-1/(H \eta)$, and we start discussing
   a case in which we allow only for a  
    homogeneous profile $\varphi$
 for the scalar field. We choose units in which at the end of inflation the scale factor is equal to one, $a_{end}=1$.   
  Consequently, in these units the scale factor is very small at the beginning of inflation, 
  $a_{in}\,=\,\exp{\left[-N_{ef} \right]}$ with $N_{ef}$ the e-fold number.   
   We 
 assume that the 
   vector does not backreact on the metric and on the inflationary dynamics; we will   critically assess these 
  hypothesis in due course.

  Let us start discussing the electromagnetic part of the action in eq \eqref{inact}.  
  We introduce 
  the combinations (the prime denotes derivative along conformal time) 
   \bea
 f^2(\varphi)&=& {\cal B}(\varphi)\,
 +\frac{2\,{\cal C}(\varphi)\,\varphi'^2}{a^2}
 \label{defF}
 \,,
 \\
 g^2(\varphi)&=&
 {\cal B}(\varphi)
   \label{defG}
 \,. 
 \eea
 We
 express the vector components as 
 \be \label{vecdec}
 A_\mu\,=\,(A_0,\,A_i+\partial_i \chi)\,,
 \ee
 with $\chi$ being the vector longitudinal
 polarization and $A_i$ the  transverse vector components  satisfying $\partial^i A_i\,=\,0$. We plug
  this decomposition into the action (\ref{inact}):
    \bea\label{redefact0or}
 S_{em}&=&\int d^3 x\,d \eta\,\left[
 \frac{f^2(\varphi)}{2}\, \left(A_i^{ '}\right)^2+\frac{g^2(\varphi)}{2}\,A_j \nabla^2 A_j
 \right]\nonumber
 \\
 &&+\int d^3 x\,d \eta\, \frac{f^2(\varphi)}{2}\,\left[2 A_0 \,\nabla^2 \chi'-A_0 \nabla^2 A_0-\chi' \nabla^2 \chi'\right]
 \,, 
 \eea
 where we neglect contributions arising from   space-dependent scalar fluctuations.
  Hence we learn that the  
  derivative contribution  to the  action (\ref{inact}), proportional
  to the function ${\cal C}$, 
 can change the effective
 sound speed of the electromagnetic field, since it makes the functions $f$ and $g$
 different. 
See  \cite{Giovannini:2013rza}  for a
  detailed analysis of or a related framework, and our Appendix  \ref{sec-pheno}.
%
   
 The fermionic action \eqref{galferm}, in this homogeneous background configuration, 
can be re-assembled as

 \bea
{\cal S}_{\psi}&=&\frac12 \int d^4 x\,\sqrt{-g}\,\Big[ i \,h^2(\varphi)\,\bar{\psi}\,\gamma^0\, \nabla_0\,\psi+
e\,h^2(\varphi)\,\bar{\psi}\,\gamma^0\,A_0\,\psi
\nonumber\\
&&\hskip2cm
+
 i \,\bar{\psi}\,\gamma^i\, \nabla_i\,\psi+e \,\bar{\psi}\,\gamma^i\,\left(A_i+\partial_i \chi\right)\,\psi+h.c.
\Big]\label{newsel}\,,
\eea
with
\bea
\label{eqforh}
{h}^2(\varphi)&=&1+k_0\, \frac{\varphi'^2}{a^2} \,.
 \eea

The essential feature  that we 
need 
  is the fact that  the coefficients in front of time and spatial derivatives of the fermion fields can be different. We will make
 use of this fact in Section \ref{sec-strong} when addressing the strong coupling problem for our scenario
  of magnetogenesis.
 Combining
 together the fermionic Lagrangian of  eq. (\ref{newsel})  with the electromagnetic Lagrangian (\ref{redefact0or}), we find 
 the following constraint equation for $A_0$
 \be\label{condA0n}
 A_0\,=\,\chi'-\frac{e\,h^2\,a^4}{f^2}\,\vec{\nabla}^{-2}\,\left(\bar{\psi}\,\gamma^0\,\psi\right)
 \,.
 \ee
Plugging this condition in the total action, we find
\bea
  S_{tot}
  &=&\int d^3 x\,d \eta\,a^4\,\left[
 \frac{f^2}{2\,a^4}\,\left(A'_i\right)^2+\frac{g^2}{2\,a^4}\,A_j \nabla^2 A_j+
i \,h^2\, \bar{\psi}\,\gamma^0\, \overleftrightarrow{\nabla}_0\,\psi 
+
 i \,\bar{\psi}\,\gamma^i\, \overleftrightarrow{\nabla}_i\,\psi+e \,\bar{\psi}\,\gamma^i\, A_i\,\psi
\right]
  \nonumber\\
 && +\int d^3 x\,d \eta\,e\,a^4\,\Big[h^2\, 
 \bar{\psi}\,\gamma^0\,\psi\,\chi'
 + \,\bar{\psi}\,\gamma^i\,\psi\,\partial_i \chi
 +\dots
  \Big]\label{stotc}
  \,, \eea
  with $\bar{\psi} \,\gamma^a \overleftrightarrow{\nabla}_a\,\psi\,\equiv \,\bar{\psi}\, \gamma^a {\nabla}_a\,\psi
  - {\nabla}_a\,\bar{\psi}\, \gamma^a \,\psi
  $, while  the dots in the second line contain 
  dimension-6, gauge invariant four fermion operators 
  (not including 
  gauge field)
%
  %
     with structure 
$\left( \bar{\psi} \psi
\right)^2
$.
 The consequences of 
 such operators
    depend on the specific  UV completion for our set-up. Indeed, it is always possible to include 
 additional
 gauge invariant  four fermion operators -- 
 weighted by an appropriate 
  function of the scalar field $\varphi$ --
  with the same structure of the ones discussed,  and 
 that compensate for their effects. Such additional operators are gauge invariant and do not involve gauge fields, hence
 do not affect the magnetogenesis scenario that we are going to develop.

The previous action \eqref{stotc} is still gauge invariant. For simplicity, we can choose a unitary gauge  $\chi=0$,  that makes 
the last line of eq (\ref{stotc}) vanishing. In this gauge, the total action to consider is then
\bea
  S_{tot}
     &=&\int d^3 x\,d \eta\,a^4\,\left[
 \frac{f^2}{2\,a^4}\,\left(A'_i\right)^2+\frac{g^2}{2\,a^4}\,A_j \nabla^2 A_j
 +
i \,h^2\, \bar{\psi}\,\gamma^0\, \overleftrightarrow{\nabla}_0\,\psi 
+
 i \,\bar{\psi}\,\gamma^i\, \overleftrightarrow{\nabla}_i\,\psi
 +e \,\bar{\psi}\,\gamma^i\, A_i\,\psi
\right]\,.
  \nonumber\\
 \label{stot}
  \eea
  We will suppose  that at the end of inflation the scalar $\varphi$ ceases to roll and stabilizes on some minimum 
  of its potential, so that $h=1$ and the quantities $f$ and $g$ become identical. 
  Such a scenario has interesting
  phenomenological consequences. 
     In Appendix \ref{sec-pheno} we  show
        that 
       it is possible to generate a scale invariant spectrum for a magnetic field without causing large backreaction
       on the inflationary energy density. In particular,  we  generalize
         the kinetically coupled model of  \cite{Ratra:1991bn} to include 
        the effect of a time dependent sound speed for the electromagnetic waves. 
       In the next section, instead, we focus on the main point of this work:
        by 
        exploiting our particular
       form of the fermionic action,  it is possible to avoid the so-called strong coupling problem for
        primordial magnetogenesis, that has been 
       first  pointed out in \cite{Demozzi:2009fu}.

\section{Avoiding  the strong coupling problem during inflation}
\label{sec-strong}

In the standard kinetically coupled scenario for magnetogenesis  introduced by Ratra in \cite{Ratra:1991bn}, it  is possible
to generate a sufficiently 
 large amplitude of  
 scale invariant  magnetic field without encountering backreaction
problems during inflation. In the simplest realizations of Ratra's scenario, one finds  
 the condition  that 
  the function $f$ appearing in eq \eqref{stot} is an increasing function of the scale
factor, $f=a^2$. See e.g. \cite{Durrer:2013pga,Subramanian:2009fu} for reviews.

  In our set-up,  we have the two independent functions $f$ and $g$ to exploit for generating  
a scale invariant  spectrum for the magnetic field.
We   show in Appendix \ref{sec-pheno} that a scale invariant magnetic field can be generated if
\be \label{secf}
f\,=\,a^{\alpha}\hskip0.5cm{\rm with }\hskip0.5 cm\,\alpha\,\ge\,2\,\,\,,
\ee
provided that
\be
g\,=\,a^\beta\hskip0.5cm{\rm with }\hskip0.5 cm\,\beta=\frac{3\alpha+4}{5}\,.
\ee
This implies that  we have more parameter space available  
 for obtaining a  scale invariant spectrum for the     magnetic field, than in the standard scenario developed
 by Ratra \cite{Ratra:1991bn} (recovered 
 for the  special 
 choice  $\alpha=\beta=2$).  
Recall that we choose units in which the scale factor starts very small at the beginning of inflation, $a_{in}\,=\,
\exp{\left[-N_{ef} \right]}$ with $N_{ef}$ the e-fold number, while 
  at the end of inflation   $a_{end}=1$. Starting  from $a_{end}$   
   the functions
$f$ and $g$ coincide since, as explained in the previous section, we make the hypothesis that the 
scalar $\varphi$ stops  rolling at the end of inflation.

\smallskip

Let us consider the consequences of these facts for the coupling between the electromagnetic field
and fermions during inflation, thinking  for definiteness to $\psi$ as the electron field. 
 The fermionic action that we investigate can be read from eq. (\ref{stot}) (where we select the gauge $\chi=0$):
 \bea
  S_{\psi}
   &=&\int d^3 x\,d \eta\,a^4\,\left[i \,h^2\, \bar{\psi}\,\gamma^0\, \overleftrightarrow{\nabla}_0\,\psi 
+
 i \,\bar{\psi}\,\gamma^i\, \overleftrightarrow{\nabla}_i\,\psi+e \, \bar{\psi}\,\gamma^i\, A_i\,\psi
\right]\,,
 \label{stot2}
  \eea
  where $h$ 
  is
\bea\label{defh}
{h}^2(\varphi)&=&1+ k_0\,\frac{\varphi'^2}{a^2} 
\,,
\eea
   In a standard scenario, without the derivative couplings 
  proportional to $k_0$ in the fermionic action  (\ref{galferm}),
  one encounters a  strong coupling problem for primordial magnetogenesis \cite{Demozzi:2009fu}.  

For setting  the stage, 
let us explicitly discuss this problem  as a special case of our discussion, by  choosing  $k_0=0$ and so (by eq \eqref{defh}) $h=1$.
 After canonically normalize the electromagnetic gauge potential (see eq. (\ref{stot})) 
  and the fermionic  electron field (see eq. \eqref{stot2}) as 
\be
A_\mu\to \frac{\hat{A}_\mu}{f}
\hskip0.5cm,
\hskip0.5cm
\psi\to\frac{ \hat \psi}{a^2 }\,,
\ee
  one 
 obtains the following effective coupling between the fields involved
 \bea \label{scint}
  {\cal L_{A\psi\psi}}&=&\frac{e}{f}\,\bar{\hat \psi}\,\gamma^i\,\hat \psi\,\hat A_i\,.
 \eea
So  
  the effective coupling 
   scales as $e/f$. But we know that  $f$ is a monotonic increasing function
 during inflation, see eq \eqref{secf}: hence    at early times  $f$ is very small. This  makes the effective coupling
$e/f$ extremely large at early   inflationary stages, thus leading
to a strong coupling regime in which the theory cannot be trusted \cite{Demozzi:2009fu}. 
  This is a general feature of conformally coupled models, although possible ways-out might  be found
  for particular, non-monotonic   profiles of the conformal functions \cite{Ferreira:2013sqa}, or
  adding helicity to the electromagnetic field \cite{Caprini:2014mja}. For other examples not involving scalar fields see
   \cite{Koivisto:2011rm}.
  
\smallskip

In our case,   including 
the  derivative couplings proportional to the quantity $k_0$,  we have the additional function $h$ at our disposal.
    The 
   canonical normalization for vector and  fermion fields now explicitly depends on $h$: 
   \be
   A_\mu\to \frac{\hat{A}_\mu}{f}
\hskip0.5cm,
\hskip0.5cm
   \psi\to \frac{\hat{\psi}}{a^2\, h}\,.
   \ee
Hence    the coupling between vector and fermion fields in eq \eqref{stot}
   is controlled by the effective Lagrangian 
   \bea
   {\cal L_{A\psi\psi}}&=& \frac{e }{f\,h^2}\,\bar{\hat \psi}\,\gamma^i\,\hat \psi\,\hat A_i
      \label{efcfv}
   \eea
   so we gain a factor $1/h^2$  with respect to eq. \eqref{scint}.
  We can now investigate situations where the time dependence of    the quantity  $1/h^2$  compensates  the
  function $f$ in the denominator of the previous formula.
      As the simplest  example, we can demand that
  \be\label{imprel}
  \frac{1}{f\,h^2}\,=\, 1\,,\hskip0.6cm 
    \, \ee
    that is, the function $h$ compensates exactly for the smallness of $f$ in the early stages of inflation.  
         Making this choice,
  eq (\ref{efcfv})
 rewrites  
    \bea
   {\cal L_{A\psi\psi}}&=&e\,
    \,\bar{\hat \psi}\,\gamma^i\,\hat \psi\,\hat A_i\,.
   \eea
     Hence 
    the effective coupling between fermion and gauge boson
    is now constant during 
     inflation.
 So it
      does not suffer from
the      
    strong coupling problem that affects the standard conformally  coupled,  $h=1$ scenario.

Using the definition for $h$ and for $f$,
 the condition \eqref{imprel} can be expressed as
\be\label{secpro}
1+ k_0\,\frac{\varphi'^2}{a^2}\,=\,a^{
 -\alpha}\hskip0.6 cm {\rm with} \hskip0.6 cm 
 \alpha\ge 2
\ee

We can think of two different ways to satisfy this condition:
\begin{itemize}
\item[-]
The first possibility is to make   the time derivative 
  $\varphi'$ of the scalar field
    {\it very large} at the beginning of  inflation, when $a_{in}\,=\,\exp{\left[
   -N_{ef}\right]}$.  In this regime, 
  the condition of the  previous formula  can be 
  re-expressed as (recall we are working in conformal time) 
  \be \label{condvp}
   k_0\,
   \left( {\partial_a\,\varphi}\right)^2\,\frac{
   a'^2}{a^2}
\,\simeq\,a^{-\alpha}
   \hskip1cm \Rightarrow \hskip1cm
    \varphi\,\simeq \,\frac{2}{\alpha\,\sqrt{k_0}\,H}\, a^{-\frac{\alpha}{2}}
  \ee  
     during  the early epoch of the inflationary quasi-de Sitter era. 
        So the scalar $\varphi$ is  in a   {\it fast-rolling} regime  during these 
      initial stages of inflation, changing with a rate depending on a parameter $\alpha\ge2$,  and
      does not   correspond to the usual slowly-rolling inflaton field.
         At the same time, we have to demand that its
      total energy density does not dangerously backreact on the geometry. 
        Possible realizations of these conditions can be found;
         an explicit one will be analyzed  in Appendix \ref{sec_expex} 
          by identifying our field $\varphi$ with   an auxiliary  
           scalar  during 
         inflation, whose kinetic  and potential energies are modulated by a suitable function of the  
        inflaton field. 
       \item[-] The second possibility is to promote the parameter $k_0$ to a  function 
       of the field profile  $\varphi$, able to satisfy eq \eqref{secpro}  without  demanding a large time derivative
       for $\varphi$.
       Following this route, one can embed this mechanism in a standard, slow-roll inflationary model   
       with $\varphi'$ small during inflation.  For example, in single field slow-roll inflation, one has the approximate
       equality
       \be
       \frac{\varphi'^2}{a^2}\,\simeq\,2\,\epsilon\,H^2
       \ee     
       with $H$ the Hubble parameter during inflation, and $\epsilon\,=\,-{H'}/(a\,H^2)<1$ is a slow-roll parameter (recall we are taking derivatives along conformal time).  
      In this case, in order to satisfy the requirement \eqref{secpro}, we can demand that the function $k_0$ has a scalar-dependent profile
       \be
       k_0\,\simeq\,\frac{a^{-\alpha}}{2\,\epsilon\,H^2}
       \ee
       during the first stage of inflation.
       \end{itemize}

   \smallskip

    We have shown that, provided that  gauge-invariant  derivative interactions are included, 
      we can avoid a strong coupling problem 
       between fermionic and electromagnetic fields during the early stages of inflation. 
 Further important requirements 
    have to be taken into account  for obtaining a  fully satisfactory    scenario of primordial  magnetogenesis.
   For example, one   should analyze in details the dynamics 
    of fluctuations of the scalar $\varphi$ and the geometry, and ensure that no additional strong coupling issues  
    emerge when coupling  such fluctuations with  the electromagnetic and
    fermionic  fields~\footnote{We thank Marco Peloso, Ricardo Ferreira and  Jonathan Ganc  for discussions on this subject.}.  
  For example, one might worry about   
  dimension-8 operators involving  fluctuations $\delta  \varphi$  of our field $\varphi$  coupled to fermions and gauge fields. Schematically such operators have the form $\partial_\mu \delta  \varphi \,
 \partial^\nu \delta  \varphi\,\left(\bar{\psi}\,\gamma^\mu\,A_\nu\,\psi \right) $, and 
 they originate from second order perturbations of  the operator weighted by $k_0$ in eq \eqref{galferm}.
  Once all  the fields  are canonically normalized, such operators
  are weighted by a factor $c_0$ that scales as  $c_0\,=\,1/(f\,E_{kin})$ during the early stages of inflation, with $E_{kin}$ the scalar kinetic energy.
    In scenarios where 
   the scalar kinetic energy does not backreact on the inflationary expansion, so to  satisfy  the inequality $E_{kin} \,\ll\, H^2 \,M_{Pl}^2$,
     $c_0$ can be well larger than $1/H^4$ during inflation (unless the scale of inflation is low), leading to a strong coupling issue. 
    Whether  or not  this is a problem depends on the explicit realization of our scenario, and on its possible UV completions including
   additional higher dimensional operators.
   For example, one
   can consider set-ups in which the scalar 
    kinetic energy is not necessarily much  smaller than the scale of inflation,  so that  
   a more careful, model dependent study is needed
    to evaluate whether such operators lead to strong coupling.
    (Since such analysis involves backreaction of scalar fields, it would nevertheless be easier to perform  with respect to a case in which it is the electromagnetic
   field that backreacts on the geometry.) Alternatively, one could consider the effect of additional gauge invariant,  higher dimensional operators coupling
   the scalar to fermions and gauge fields; one 
    example are  dimension-12 operators of the schematic form $(\partial \varphi)^2\,\left(\bar \psi\,\gamma\,A\,\psi \right)$, 
    that once expanded in terms of scalar perturbations
     contain (among others) operators of the right structure to compensate for the 
   effects of the aforementioned ones.
   
   Of course, a detailed analysis of these and other issues will be  important 
   to really understand whether it is possible to build explicit models for primordial magnetogenesis along these lines,  that are  
   under  control  when examined  order by order in perturbation theory. 
   The 
     general  proposal discussed  in this paper, if on the right track,  might   
      point towards set-ups  with a non-trivial  structure, possibly
       determined by some symmetry principle, in which the role of higher dimensional derivative operators   is crucial in defining a viable magnetogenesis model  during inflation.


\section{Outlook}

  We proposed  a framework    
    for building gauge invariant   models of primordial magnetogenesis,  that are 
 potentially free from    strong coupling problems. 
     The main idea    is   to derivatively couple 
       electromagnetic and fermion fields to a scalar field $\varphi$
    during inflation. This scalar  is not necessarily the inflaton. 
         Gauge-invariant derivative scalar-vector couplings, although explored in modified gravity scenarios,
     to the best of our knowledge have not been investigated in the context of primordial
    magnetogenesis. 
     We show that their inclusion induces a non-trivial sound speed for  electromagnetic
    and fermionic modes. Moreover, and most importantly, derivative couplings 
        give additional freedom 
      to control the time-dependence of the electromagnetic coupling
      constant during inflation,  by making this quantity depending 
        also on the {\it time derivative} of the scalar background
       profile.   This fact allows us to find conditions to avoid the serious strong coupling issue that, as
        first pointed out in  \cite{Demozzi:2009fu}, 
        affects the simplest realizations
       of magnetogenesis scenarios. 
    
    Besides potentially avoiding the  strong coupling problem,
     our scenario
     allows  us to generate a large scale magnetic field of sufficient amplitude, that does not suffer from 
     large backreaction issues.
      We do not need to rely on explicit inflationary  models for developing  our ideas, that 
might   
 be applied 
to different cases.
    On the other hand,   specific  
 conditions   have to be imposed on
  our system of
   a scalar  derivatively coupled  to the electromagnetic and
 fermionic actions. A first realization
  of our mechanism requires   that the homogenous profile for 
  scalar field has a large time derivative during  the early stages of the inflationary era. We shown in an explicit
  concrete example how to satisfy such conditions in a consistent way. 
A second realization instead imposes  that some of  the functions 
   of the scalar field, that multiply our derivative operators, vary considerably during inflation. 
  
 \smallskip

 It would  be interesting to    apply  more broadly 
  these ideas  to  specific models of inflation, to  study at which extent  
   the requirements
 of obtaining  theoretically convincing   magnetogenesis set-ups 
   can constrain a given inflationary model.
        When analyzing explicit  inflationary models,  
       it will be important to 
          carefully study 
       the dynamics of scalar  perturbations,  and how the   derivative couplings to the electromagnetic  field 
         can affect their behaviour. 
  Indeed,  in recent years there have been  many investigations 
       of possible observational consequences of primordial magnetogenesis
      for the production of CMB anisotropies and non-Gaussianities (see e.g. \cite{Bonvin:2013tba,Caprini:2009vk,
      Jain:2012vm,Nurmi:2013gpa}), and more in general  on constraints on magnetogenesis
      scenarios from CMB observations \cite{Fujita:2014sna,Fujita:2012rb,Fujita:2013qxa,Kobayashi:2014zza}.
        These studies have been generally
      done taking the kinetically  coupled magnetogenesis model of \cite{Ratra:1991bn} 
      as reference scenario. It would  be
      interesting to generalize these results to the set-up presented in this work   including a time-dependent  sound speed
      for the electromagnetic potential,  and taking into account the full dynamics of the  fluctuations
      for the fields involved.

\acknowledgments
It is a pleasure to thank Marco Peloso (especially) and Lorenzo Sorbo for their  initial  collaboration on this work, 
  for many useful discussions and constructive criticisms, and for comments
  on a draft. Most of the results contained in Appendix \ref{sec-generating} have been obtained in collaboration with them.
Also,
 thanks to Takeshi Kobayashi, Vincent Vennin, and Ivonne Zavala for
   interesting  discussions, and thanks to Ricardo Ferreira and Jonathan Ganc for useful 
   comments on a first version of this manuscript and for their observations raised in \cite{Ferreira:2014mwa}.
     GT is supported by an STFC Advanced Fellowship ST/H005498/1. 
 He also acknowledges  the Munich Institute for Astro- and Particle Physics (MIAPP) of the DFG cluster of excellence "Origin and Structure of the Universe", and  the Yukawa Institute, Kyoto, 
 for  financially  support    visits 
  during which this work has been developed.

\begin{appendix}

  \section{Phenomenology of this  scenario}\label{sec-pheno}

\subsection{Generating a scale invariant spectrum for the   magnetic field}\label{sec-generating}

 We now   investigate
 how 
  our set-up 
   can lead to the production  of
   a homogeneous magnetic field  with scale invariant spectrum.
    A
  detailed analysis of a  framework  related to ours has been recently carried on in 
    \cite{Giovannini:2013rza}.
   %
       To start with,  
  we focus on the consequences of the  scalar-vector  
   action  \eqref{inact}, neglecting
the effects of fermion production during inflation, that we
assume to be negligible \footnote{
Let us  point out that the derivative couplings
we considered are not the most general ones, and others could be examined, 
 for example motivated by Galileon symmetries
  \cite{Deffayet:2010zh,Tasinato:2013oja}. On the other hand, for illustrative purposes,  the minimal structure of our 
    actions (\ref{inact}) and \eqref{galferm} will be sufficient.}.

Our purposes is to  study cosmology in a conformally flat de Sitter background (see eq \eqref{cfmet}).  
 Hence 
  we focus on the  scalar-vector  
part of the   action 
 \eqref{stot}, that is
 \be\label{redefact2}
 S_{EM}\,=\,\int d^3 x\,d \eta\,\left[
 \frac{f^2(\varphi)}{2}\,\left(A'_i\right)^2+\frac{g^2(\varphi)}{2}\,A_j \nabla^2 A_j
 \right]\,.
 \ee
    We relate the derivatives of the gauge potential to the electric and magnetic field via the following formulae
\begin{eqnarray}
F_{0i} &  = & a^2 E_i \,, \nonumber\\
F_{ij} &  = & a^2 \epsilon_{ijk} B_k \,,
\end{eqnarray}
that allow us to rewrite our action (\ref{redefact2})  as 
\begin{equation}\label{newaction}
S = \frac{1}{4} \int d^4 x \sqrt{-g} \, {\cal L} = \int d^4 x \left[  \frac{f^2}{ 4 }  F_{0i} F_{0i} -  \frac{g^2}{4 } F_{ij} F_{ij}  \right]
\,.
\end{equation}
This action differs from the usual `kinetic coupled' model \cite{Ratra:1991bn} since our construction allows us for  a 
 time-dependent   sound speed for
the electromagnetic potential  when $f\neq g$. 
    We have learned in Section 
   \ref{sec-vesc} 
   that (neglecting fermions) the constraint equation eliminates the longitudinal polarization $\chi$
   of the vector, hence
 $\partial_i A_i = 0$.
  We therefore decompose
\bea
\vec{A} & = & 
\sum_{\lambda=\pm} \int \frac{d^3 k}{\left( 2 \pi \right)^{3/2}} 
\vec {\bf  e}_\lambda ( \vec{k} )
 { e}^{i \vec k \cdot \vec x} \, \frac{{\hat V}_\lambda ( k )}{f}\,, \nonumber\\
& =  &
\sum_{\lambda=\pm} \int \frac{d^3 k}{\left( 2 \pi \right)^{3/2}} 
\vec {\bf e}_\lambda ( \vec{k} )
 { e}^{i \vec k \cdot \vec x} \, 
\left[  \, a_\lambda ( \vec{k} ) \, \frac{V_\lambda ( k )}{f}  + 
 a_\lambda^\dagger ( - \vec{k} ) \, \frac{V_\lambda^*  ( k )  }{f}     \right] \;\;\;,\;\;\;
\label{A-deco}
\eea
where the circular polarization operators satisfy $\vec{k}\cdot \vec {\bf e}_{\pm} ( \vec{k} ) = 0$, $\vec{k} \times 
\vec {\bf e}_{\pm} ( \vec{k} ) = \mp i k \,\,\vec {\bf e}_{\pm} ( \vec{k} )$, $\vec {\bf e}_\pm ( \vec{-k} ) = \vec {\bf e}_\pm^* ( \vec{k} )$, and are  normalized according to $\vec {\bf e}_\lambda^* ( \vec{k} ) \cdot \vec {\bf  e}_{\lambda'} ( \vec{k} ) = \delta_{\lambda \lambda'}$. The annihilation and creation operators appearing
 in \eqref{A-deco} 
satisfy the commutation relations
\be\left[ a_\lambda ( \vec{k} ) ,\, a_{\lambda'}^\dagger (  \vec{k'} ) \right] = \delta_{\lambda \lambda'} \, \delta^{(3)} \left( \vec{k} - \vec{k}' \right)\,.\ee

\smallskip

In  terms  of the Fourier quantum field ${\hat V}_\lambda$,   our action (\ref{newaction}) results
\bea
S & = & \frac{1}{2} \sum_\lambda \int d t\, d^3 k\, \left[ f^2 \, \left\vert \left( \frac{{\hat V}_\lambda \left( k \right)}{f} \right)' \right\vert^2 -  \frac{g^2}{f^2} k^2 \, \left\vert {\hat V}_\lambda \left( k \right) \right\vert^2 \right] \,,\nonumber\\
& \equiv &   \frac{1}{2} \sum_\lambda \int d t \,d^3 k \,\left[ \vert {\hat V}_\lambda' \vert^2 - \left( \frac{g^2}{f^2} k^2 - \frac{f''}{f} \right) \vert
 {\hat V}_\lambda' \vert^2  \right]
\,,
\eea
   and the equation of motion for the mode function is
\be
V_\lambda'' + \left( \frac{g^2}{f^2} k^2 - \frac{f''}{f} \right) V_\lambda = 0\,.
\ee
This variable $V_\lambda$ is canonically normalized. 
We decompose the electric and magnetic fields as
\bea 
\vec{ E} & = & \sum_{\lambda=\pm} \int \frac{d^3 k}{\left( 2 \pi \right)^{3/2}} \left[
\vec {\bf e}_\lambda \left( \vec{k} \right)
 { e}^{i \vec k \cdot \vec x} \, 
  \, a_\lambda \left( \vec{k} \right) \,  E_\lambda \left( k \right) + { h.c. }\right] \,,\nonumber\\
\vec{ B} & = & \sum_{\lambda=\pm} \int \frac{d^3 k}{\left( 2 \pi \right)^{3/2}} \left[
\vec {\bf e}_\lambda \left( \vec{k} \right)
 { e}^{i \vec k \cdot \vec x} \, 
  \, a_\lambda \left( \vec{k} \right) \,  B_\lambda \left( k \right) + { h.c. } \right]\,,
\eea
and their mode functions are  related to $V_\lambda$ by ($\lambda\,=\,\pm$) 
\be
 E_\lambda = - \frac{1}{a^2} \left( \frac{V_\lambda}{f} \right)' \;\;\;,\;\;\;
 B_\lambda = \frac{\lambda \,k}{a^2 }   \frac{ V_\lambda }{ f }     \,.
\ee

\smallskip

 We proceed computing the energy density associated with this system. We have
\begin{equation}   
\rho =       - T^0_0  =     \frac{f^2}{2 a^4} F_{0i} F_{0i} + \frac{g^2}{4 a^4} F_{ij} F_{ij}  
= \frac{1}{2}    \left[ f^2 E^2 + g^2 B^2 \right]\,,
\end{equation}  
and so
\bea
\langle \rho \rangle &=& \sum_\lambda \int \frac{d^3 k}{\left( 2 \pi \right)^3} \left\{ \frac{f^2}{2} \Big\vert E_\lambda \left( k \right) \vert^2 
+  \frac{g^2}{2} \Big\vert B_\lambda \left( k \right) \vert^2 \right\}
\\
 & \equiv & \langle \rho_E \rangle +  \langle \rho_B \rangle
\,.
\eea
 The total energy is given by a sum  of magnetic and electric contributions, with energy
densities
\begin{eqnarray}
\frac{d \langle  \rho_E \rangle }{d {\rm ln } k } & = & \sum_\lambda \frac{k^3}{4 \pi^2} f^2 \vert E_\lambda \left( k \right) \vert^2 
= \sum_\lambda \frac{k^3}{4 \pi^2 a^4} f^2 \Bigg\vert \left( \frac{V_\lambda}{f} \right)' \Bigg\vert^2 \,,\nonumber\\
\frac{d \langle  \rho_B \rangle }{d {\rm ln } k } & = &   \sum_\lambda \frac{k^3}{4 \pi^2} g^2 \vert B_\lambda \left( k \right) \vert^2 
= \sum_\lambda \frac{k^5}{4 \pi^2 a^4} \frac{g^2}{f^2} \Big\vert  V_\lambda  \Big\vert^2 
\label{expend1}\,.
\end{eqnarray}

\smallskip

For definiteness, 
 we now  adopt a power-law Ansatz for the time-dependence of the functions $f$ and $g$:
\begin{equation}\label{deffg}
f(\eta)\,=\,a^{\alpha}\,=\,\left(-H \eta \right)^{-\alpha}\,,\qquad g(\eta)\,=\,a^{\beta}\,=\,\left(-H \eta \right)^{-\beta}\,,
\end{equation}
and we set $\eta_{end}\,=\,-1/H$ as the time at which inflation ends, when $f=g=a_{end}=1$. 
This choice leads to the following equation for $V_\lambda$
\begin{equation}\label{v-eq}
V_\lambda''  + \left[ \left(-H\,\eta\right)^{ 2\alpha-2 \beta}\,k^2-\frac{\alpha(\alpha+1)}{\eta^2}\right] V_{\lambda} = 0
\,.
\end{equation}

As   mentioned above, 
 the sound speed (more appropriately, the speed of light) of photons  at small scales  is given  by $c_{eff}^2 ={g^2}/{f^2}$, so that
\be
\label{sspeed}c_{eff}\,=\,\left(-H\,\tau\right)^{ \alpha-\beta}\,=\,a^{\beta-\alpha}\,\,.
\ee 
  To avoid superluminal propagation (which, in turns, may create problems for
    finding an  UV completion for the model \cite{Adams:2006sv}),  we  require that  $g^2 \le f^2$.
            This implies that the speed of light starts very small at the beginning of inflation (i.e. much smaller than the speed 
      of gravitons) and approaches the value $c_{eff}=1$ at the end of inflation.
          Therefore, we impose that
                     $  \beta > \alpha$.

We also require
   that the part proportional to $k$ in eq (\ref{v-eq}) dominates the coefficient of $V_\lambda$ at early times, {\em {i.e.}} that
 \begin{eqnarray}
 \alpha -\beta +1 > 0 \,.
  \end{eqnarray}

This condition guarantees the validity of the adiabatic approximation at early times, so that we can assume that modes of fixed momentum $k$ are initially in the adiabatic vacuum state.
  The solution of eq.~(\ref{v-eq}) that reduces to positive frequency modes at early times is
(up to an unphysical phase)

\vskip0.5cm

\begin{eqnarray}
V_\lambda & =& \frac{i}{2\sqrt{1+\alpha- \beta}} \, \sqrt{\frac{\pi}{a\,H} } \, H_{\frac{1+2 \alpha}{2\left(1+\alpha-\beta\right)}}^{(1)} \left( 
 \frac{k}{H}  \frac{ a^{\beta-\alpha-1} }{\alpha-\beta+1} \right) 
 \nonumber\\
 \alpha >- \frac{1}{2}\,\,:
    \hskip3.3cm&&  \label{geso1}
 \\ \nonumber
f \left( \frac{V_\lambda}{f} \right)' &=& - i \frac{k \sqrt{\pi}}{2 \sqrt{H} \sqrt{1+\alpha-\beta}} a^{\beta-\alpha 
-
\frac{1}{2}}
H^{(1)}_{-\frac{1-2\beta}{2 \left( 1 + \alpha -
  \beta \right)}} \left( \frac{k}{H}  \frac{a^{\beta-\alpha-1}  }{\alpha-\beta+1} \right)
\end{eqnarray}
and
\begin{eqnarray}
V_\lambda &=& \frac{i}{2 \sqrt{1+\alpha-\beta}} \,  \sqrt{\frac{\pi}{a\,H} } \, H_{-\frac{2 \alpha +1 }{2\left(1+\alpha-\beta\right)}}^{(1)} \left( 
 \frac{k}{H}  \frac{ a^{\beta-\alpha-1} }{\alpha-\beta+1} \right) 
  \nonumber\\
 \alpha <- \frac{1}{2}\,\,:
    \hskip3.3cm&& \label{geso2}
 \\ \nonumber
f \left( \frac{V_\lambda}{f} \right)' &=&  i \frac{k \sqrt{\pi}}{2 \sqrt{H} \sqrt{1+\alpha-\beta}} a^{\beta-\alpha
-
\frac{1}{2}}
H^{(1)}_{\frac{1-2\beta}{2 \left( 1 + \alpha - \beta \right)}} \left( \frac{k}{H}  \frac{ a^{\beta-\alpha-1}}{\alpha-\beta+1} \right) 
\end{eqnarray}
with $H^{(1)}_\nu(x)$ denoting the Hankel function of the first kind. 

\vskip0.5cm

These solutions have the correct asymptotic early time limits, and the arbitrary phase is chosen so that the modes 
 associated with 
 $V_\lambda$ are real and positive at late times.

Let us focus on the simplest situation of particular interest  for generating a large-scalar magnetic field: the case of a scale invariant magnetic energy density. Plugging eqs (\ref{geso1}, \ref{geso2})
 into eq \eqref{expend1}, we find the following 
 condition 
 
%
\begin{eqnarray}
\frac{d \langle  \rho_B \rangle }{d {\rm ln } k } \Big\vert_{\rm late}   =   {\rm scale\,\, independent} & \;\; {\rm for  } \;\; & \alpha >- \frac{1}{2} \;\; {\rm and } \;\; \beta = \frac{3 \alpha + 4}{5}\,, \nonumber\\
& {\rm or } &  \alpha <- \frac{1}{2} \;\; {\rm and } \;\; \beta = \frac{7 \alpha + 6}{5} \label{sinreq}\,.
\end{eqnarray}
Fixing the parameter $\beta$ as in the previous formulae  allows us to greatly simplify the expressions in equations (\ref{geso1}, \ref{geso2}), since the index of the Hankel function acquires the value $5/2$, 
 and we get the exact solutions (valid at all times)
\begin{eqnarray}
\alpha > - \frac{1}{2}
& : & 
V_\lambda = \frac{3 \left( 1 + 2 \alpha \right)^2 H^2 a^{\alpha}}{25 \sqrt{2} k^{5/2} } \left( 1 - i z - \frac{z^2}{3} \right) {\rm e}^{i z} \,,\nonumber\\
& & 
f \left( \frac{V_\lambda}{f} \right)' =  \frac{-  \left( 1 + 2 \alpha \right) H a^{\frac{3+\alpha}{5}} }{5 \sqrt{2} k^{1/2} } \left( 1 - i z  \right) {\rm e}^{i z} \label{amfr}
 \,,\end{eqnarray}
and
\begin{eqnarray}
\alpha < - \frac{1}{2}
& : & 
V_\lambda = \frac{3 \left(   2 \alpha + 1  \right)^2 H^2 a^{-\alpha-1}}{25 \sqrt{2} k^{5/2} } \left( 1 - i z - \frac{z^2}{3} \right) {\rm e}^{i z} \,,\nonumber\\
& & 
f \left( \frac{V_\lambda}{f} \right)' =  \frac{3 \left( 2 \alpha +1 \right)^3 H^3 a^{-\alpha} }{25 \sqrt{2} k^{5/2} } \left( 1 - i z - \frac{2 z^2}{5} + \frac{i z^3}{15}   \right) {\rm e}^{i z} 
\,,
 \end{eqnarray}
where $z \equiv \frac{5}{\vert 1 +2 \alpha \vert} \frac{k}{H} a^{-\frac{\vert1+2\alpha\vert}{5}}$. 
 Notice  that, for this choices of $\beta$, we have
\be 
1+\alpha - \beta = \frac{\vert1+2\alpha\vert}{5}
\ee
 in both regimes, and so we automatically satisfy the  $ 1+\alpha - \beta   > 0 $ condition.

\bigskip

From these exact, all-time solutions, we get the late time energy densities (summing  over the two photon  polarizations)
\begin{eqnarray}
\alpha >- \frac{1}{2} & : & \;\;\;  
\frac{d \langle  \rho_B \rangle }{d {\rm ln } k } \Big\vert_{\rm late}   =  \frac{9 \left( 1 + 2 \alpha \right)^4 H^4}{2500 \pi^2} a^{-\frac{6 \left( 2 - \alpha \right)}{5}} \;\;,\;\;
\frac{d \langle  \rho_E \rangle }{d {\rm ln } k } \Big\vert_{\rm late}   =  \frac{ \left( 1 + 2 \alpha \right)^2 H^4}{100 \pi^2} 
\left( \frac{k}{H} \right)^2 a^{-\frac{2 \left( 7 - \alpha \right)}{5}} \,,\nonumber\\
\alpha <- \frac{1}{2} & : & \;\;\;  
\frac{d \langle  \rho_B \rangle }{d {\rm ln } k } \Big\vert_{\rm late}   =  \frac{9 \left(   2 \alpha + 1 \right)^4 H^4}{2500 \pi^2} a^{\frac{-6 \left(  \alpha+3 \right)}{5}} \;\;,\;\;
\frac{d \langle  \rho_E \rangle }{d {\rm ln } k } \Big\vert_{\rm late}   =  \frac{ 9 \left(   2 \alpha + 1 \right)^6 H^4}{2500 \pi^2} \left( \frac{H}{k} \right)^2 a^{-2\alpha-4} \,.\nonumber\\
\label{scaleinv}
\end{eqnarray}

From the previous expressions, one learns that
  the $\alpha <- \frac{1}{2} $ case leads to  a too large electric field energy produced
  during inflation.
   Indeed, ${d \langle  \rho_E \rangle }/{d {\rm ln } k } $
    is divergent in the infrared limit $k\to0$, signalling a strong backreaction of the electric field on the inflating background.
     Therefore we disregard this choice. Notice instead  that  in the $\alpha >- \frac{1}{2}$ regime 
      we have 
\begin{eqnarray}
V_\lambda V_\lambda^{*'} - { \rm h. c. } & = & 1\,, \nonumber\\
V_\lambda V_\lambda^{*'} + { \rm h. c. } & = &   \frac{9 \alpha \left( 1 + 2 \alpha \right)^4}{625} \left( \frac{H}{k} a^{\frac{1+2\alpha}{5}} \right)^5 - \frac{3 \left( 1 +2 \alpha \right)^2 \left( 1 - 3 \alpha \right)}{125}  \left( \frac{H}{k} a^{\frac{1+2\alpha}{5}} \right)^3 - \frac{2-\alpha}{5} \left( \frac{H}{k} {a^{\frac{1+2\alpha}{5}}} \right)  
\,. 
\nonumber\\
\end{eqnarray}
Since ${a}$ strongly increases during inflation, and since $1+2 \alpha > 0$, the mode function is strongly increasing, and there is classicalization for all $\alpha >- \frac{1}{2}$: the energy associated to the classical field is finite.

\smallskip

So we focus our attention to a scenario
 with  a scale-invariant  magnetic energy density,  with parameters chosen in the  range 
 \be \label{regime}\alpha\,>\,-\frac12 \hskip0.5cm {\rm and} \hskip0.5cm \beta\,=\,\frac{3\alpha+4}{5}
 \,.\ee
The standard case of unit sound speed, $\alpha=\beta$, is obtained by choosing $\alpha=2$. In our case
we have freedom to choose any preferred value $\alpha>-1/2$, although
the parameter $\beta$ has then to be tuned accordingly to
(\ref{regime}).

\smallskip

Let us calculate  
the amplitude of magnetic field  towards the end of inflation, that occurs at $a=a_{f}=1$. 
 This quantity
 can be estimated by the general formula
(see e.g. \cite{Demozzi:2009fu})
\bea\label{findb}
\delta_B^2&=&
\frac{d\,\langle \rho_B\rangle}{d\,\ln k} {\Big|_{ a=1}}
\\
&=&
\frac{9\,(1+2\alpha)^4\,H^4}{2500\,\pi^2}\,.
\eea
So the spectrum of the magnetic field is 
 scale invariant. 
 
 For values of $\alpha$ of order unity,
 one obtains a magnetic field of amplitude $\delta_B\,\simeq\, \left(H/M_{Pl} \right)^2\,10^{58}\, G$ when inflation
ends. After the end of inflation, $\delta_B^2$ decays as $1/a^4$, exactly as radiation: it is straightforward
  to estimate the amplitude
of magnetic field today in the scale invariant case
 \cite{Demozzi:2009fu,Durrer:2013pga,Subramanian:2009fu}
\be\label{condB}
\delta_B\,\simeq\,5\,\times\,10^{-10}\,\left( \frac{H}{10^{-5}\,M_{Pl}}\right)\,G\,.
\ee 
As discussed in the introduction,
observations require an amplitude of at least $10^{-15} G$ at intergalactic scales: these values are not difficult
to obtain with a Hubble scale during inflation  larger than  $H\ge 10^{-10} M_{Pl}$, corresponding to a scale
of inflation $E_{inf}\ge 10^{-5}\,M_{Pl}$ (with $E_{inf}^4\,\sim\,3 H^2\,M^2_{Pl}$). 

\smallskip

Hence,  we shown  that allowing for derivative couplings between the scalar clock and the electromagnetic
action one can enrich the phenomenology of the conformally coupled scenario \cite{Ratra:1991bn}. 
It would be interesting to  further study phenomenological ramifications of this scenario. For example
 investigating 
 how the electromagnetic
field affects the curvature perturbation \cite{Fujita:2014sna,Fujita:2012rb}. Or examining 
 the 
 structure of correlation  involving  curvature perturbation and the 
 electromagnetic field, that have been analyzed in the conformally coupled case (see e.g. \cite{Bonvin:2013tba,Caprini:2009vk,Jain:2012vm,Nurmi:2013gpa,Fujita:2013qxa,Kobayashi:2014zza}). Or studying whether new derivative interactions generalizing the ones we considered here can have an impact on the evolution
 of the magnetic field during radiation and matter dominated era \cite{Jedamzik:1996wp}.
  It is expected that the non-trivial
 sound speed for the electromagnetic field plays an important role in characterizing the phenomenology of this set-up.
 We leave these interesting topics for future work, and focus now on ensuring that our scenario  avoids 
  to produce too much electromagnetic energy during inflation.

\subsection{Conditions to avoid large backreaction from the electromagnetic field}
\label{sec-backreac}

  Inflation is a period of quasi-exponential
cosmological expansion, that lasts for $N_{tot}=\ln{\left(\frac{a_f}{a_i}\right)}$ e-folds, where $a_{i}$ and $a_f$ are 
the values of the scale factors at the  beginning and end of inflation. 
 In order
to solve the basic problems of standard Big Bang cosmology, and generate a scale invariant
 spectrum of curvature perturbations  at large scales, one finds that 
 $N_{tot}\ge50$ \cite{Liddle:2000cg}. 
   To ensure that the electromagnetic field 
does  not dangerously  backreact  on the quasi-de Sitter inflationary expansion,  we require that the energy stored in the electric and magnetic fields is smaller than
the inflationary energy density 
  $\rho_{inf}\,=\,\left( 3\,H^2\,M_{Pl}^2\right)^{1/4}$ during inflation. 

 In our  set-up, the electromagnetic potential
  has a time-dependent  sound speed. Hence, each electromagnetic mode,  characterized by momentum $k$,   typically leaves the horizon at a different time with respect to
 corresponding mode in the  scalar inflationary  fluctuations. (A similar behaviour was first pointed out in \cite{Easson:2007dh} 
in a different context of a two-scalar inflationary set-up.)  At a given
value $a$ of the scale factor, the  electromagnetic modes that
leave the horizon have  comoving momentum $k= a H/c_s$.  During inflation,
 the electromagnetic sound speed tipically  scales  rapidly with the scale factor (since $c_s\,=\,a^{\alpha-\beta}$). These
 facts imply that in order to generate a coherent magnetic field  with scale invariant spectrum at scales of the 
 order of present-day horizon, we have to satisfy the inequality 
  \bea
 \int_{\frac{a_i\,H}{c_s(a_i)}}^{\frac{a_f\,H}{c_s(a_f)}}\,\frac{d k}{k}\,&\ge&\,50\,,
 \eea
 that translates to
\bea
 {\left(1+\alpha-\beta\right)}\,N_{tot}&\ge&\,50\,.
 \eea
 Interestingly, if $\left(1+\alpha-\beta\right)\le1$, we need more than $50$ e-folds of inflation to 
 generate a coherent scale invariant spectrum
  for the magnetic field up to very large scales. 
 Let us focus again on the case of scale-invariant magnetic fields in the range  $\alpha>-1/2$
 and $\beta\,=\,(3\alpha+4)/5$
 as dictated by eq. (\ref{regime}).
 The total
 energies stored in the electric and magnetic fields after $N_{ef}$ e-folds since the beginning of inflation (of course $N_{ef}\le N_{tot}$) are given
 by the following integrals, performed over the classical modes that crossed the horizon during the epoch of interest.
 For the total electric energy, using eqs (\ref{scaleinv}) we obtain

 \bea
\rho_E&=&\int_{a_i\,H/ c_s(a_i)}^{a\,H/ c_s(a)}\,\frac{d k}{k}\,
\frac{d\,\langle \rho_E\rangle}{d\,\ln k}
\nonumber\,,
\\
&=&\int_{a_i^{\frac{1+2\alpha}{5}} H}^{a^{\frac{1+2\alpha}{5}} H}\,\frac{d k}{k}\,
\frac{d\,\langle \rho_E\rangle}{d\,\ln k}
\nonumber\,,
\\&\simeq&\frac{(1+2\alpha)^2}{200\,\pi^2}\,H^4\,\exp{\left[\frac{6(2-\alpha)}{5}
\,\left(N_{tot}-N_{ef}\right)
\right]}\,,
\eea
where in the last equality we only wrote the dominant contribution to the integral in the range  $\alpha>-1/2$, and recall that $N_{ef}$ corresponds to the number
of e-folds since the beginning of inflation. An analogous calculation gives the   
total energy stored in the
 magnetic 
field:
\bea
\rho_B&=&\int_{a_i^{1+\alpha-\beta} H}^{a^{1+\alpha-\beta} H}\,\frac{d k}{k}\,
\frac{d\,\langle \rho_B\rangle}{d\,\ln k}\nonumber\,,
\\
&=& \frac{9\,N_{ef}\,(1+2\alpha)^5}{12500\,\pi^2}\,H^4\,\exp{\left[\frac{6(2-\alpha)}{5}
\,\left(N_{tot}-N_{ef}\right)
\right]}\,.
\eea
In both cases, we have an exponential dependence on the number of efolds: given that $\left(N_{tot}-N_{eff}\right)>0$,
to avoid large backreaction we need to impose that the coefficient of $\left(N_{tot}-N_{eff}\right)$ 
 in the exponent is negative, leading to the requirement
\be\label{condal}
\alpha\ge2\,\,.
\ee
We  impose this
condition   on this work. 
    At this point, we need to avoid large backreaction   towards the end of inflation, when $N_{eff}\simeq N_{tot}$
 and the energy stored in the electromagnetic field is maximal. The requirement that $\rho_{B,\,E}$ are 
   less than the   inflationary energy density $\rho_{inf}$  imposes the following condition
\be
\frac{9\,N_{tot}\,(1+2\alpha)^5}{12500\,\pi^2}\,H^4\lesssim 3 H^2 M_{Pl}^2\hskip0.5cm\Rightarrow
\hskip0.5cm H\,\lesssim\,\frac{100}{\sqrt{N_{tot}} (1+2\alpha)^2}\,M_{Pl}\,\,.
\ee
This requirement  can be easily satisfied for   relevant values of the Hubble parameter, as the ones discussed after
eq. (\ref{condB}).

    \section{A  fast rolling scalar field with no large backreaction during inflation}\label{sec_expex}
    
    In Section \ref{sec-strong}, 
     we learned  that, by including derivative
    couplings of a scalar field $\varphi$ to fermions, it is possible to avoid the strong coupling problem
     \cite{Demozzi:2009fu}
     for  scenarios of primordial magnetogenesis based on Ratra's idea of coupling a scalar to 
     electromagnetism during inflation. 
      The simplest realization of our set-up requires 
     that the  {\it velocity} $\varphi'$ of the scalar field is  large  during the
     early epochs of 
     inflationary era, so that the scalar is in a fast-roll regime at this stage.
   With eq \eqref{condvp}  we  established the following requirement for the homogeneous part of the scalar profile 
     \be \label{condvp2}
    \varphi\,\propto \,a^{-\frac{\alpha}{2}} \hskip0.7cm,\hskip0.7cm  \alpha\ge 2
  \ee  
    with a constant of proportionality that depends on the parameters of the model.
    
    \smallskip
    
    In this sense, then, we 
    are moving the strong coupling issue for magnetogenesis to the challenging problem of finding a scenario
    where the time-derivative of a scalar field is  large during the inflationary quasi-de Sitter
    expansion.  
    At first sight,
    this condition seems hard to satisfy, since in conventional  inflationary models scalar fields {\it slowly roll} during the inflationary
    epoch,  in order to avoid 
     that its
      kinetic energy spoils the inflationary dynamics.  On the other hand, this requirement 
       is not strictly necessary: we review
     here a scenario,  elaborated in 
      \cite{Peloso:2014oza}, 
       where a scalar field can acquire a large velocity during inflation, yet avoiding  strong backreaction problems.

       We assume  that $\varphi$ is {\it not} the inflaton; the inflaton is  denoted with $\Phi$
       and  has its own dynamics that we do not need to  discuss. The
          action for our scalar $\varphi$ is given by
       \be\label{Snoinfl}
       {\cal S}_{\varphi}\,=\,\int d^4 x\,\sqrt{-g}\,{\cal N}^2\left[\Phi(\eta)\right]\,\left[ -\frac12\,\partial_\mu \varphi \partial^\mu
       \varphi-\frac{m^2}{2}\,\varphi^2 \right]
       \ee
     The function ${\cal N}\left[\Phi(\eta)\right]$ depends  on  time since it is a function of the inflaton field $\Phi$ that
     acts as clock. We still assume   that the  metric is well approximated by a 
      conformally flat 
     de Sitter   space during inflation,   with Hubble parameter $H$.      
 By selecting a suitable inflaton homogeneous profile $\Phi(\eta)$,    
     we  assume that the function   ${\cal N}(\eta)$
       is proportional to a power of the de Sitter scale factor 
     \be
     {\cal N}\,=\,a^{\gamma}\,,
     \ee
     with  $\gamma$ a constant parameter that we choose  to be positive, 
     to avoid strong coupling problems in the scalar sector during inflation \cite{Peloso:2014oza}. 
      We make the hypothesis     
       that the scalar $\varphi$ does not interfere with the inflationary
     dynamics   characterized by an (almost) constant Hubble parameter $H$, so that the profile of ${\cal N}$ does not change during inflation.   
     It is straightforward to solve the homogeneous equation of motion for $\varphi$
     in a de Sitter geometry, obtaining    the following  
     solution
      \be\label{ldconf}
     \varphi\,=\,\varphi_0\,a^{-\left(\frac32+\gamma\right)
     \,\left[1- \sqrt{1-\frac{4 m^2}{ H^2\,(3+2 \gamma)^2}} \right]}
     \ee
      for a constant parameter 
        $\varphi_0$, that we can imagine to select with an appropriate choice of initial conditions.
       By tuning
        the quantities $\gamma$ and $m/H$ we can obtain the preferred exponent  of the scale factor in the previous solution, to match the condition \eqref{condvp2} with the preferred value for $\alpha$.
        
       The  energy density associated with the configuration \eqref{ldconf}  
       is
       \be
       \rho\,=\,-T_0^{\,\,0}\,\,=\,\frac{\varphi_0^2\,H^2}{4}\,\left(3+2 \gamma\right)^2 \,
       \left[1- \sqrt{1-\frac{4 m^2}{ H^2\,(3+2 \gamma)^2}} \right]\,a^{3 \left[ \sqrt{\left(1+\frac{2 \gamma}{3}\right)^2-\frac{4 m^2}{9 H^2}} -1\right]}
  \,.     \ee
     In order to avoid  a large backreaction of the scalar energy density at the early stages of inflation, when $a\simeq e^{-N_{ef}}$, we demand that the power of the scale factor in the previous expression is larger or equal than zero, requiring
     \be\label{inega}
\left(1 +\frac23    \gamma\right)\ge   \sqrt{1+ \frac{4 m^2}{9 H^2}}\,.
     \ee
       This  ensures us 
        that our scalar configuration does not strongly backreact on the geometry during inflation, for  values of
        $\varphi_0$  smaller than $M_{Pl}$. 
        
        \smallskip
        
        This particular  realization  
        of a scalar action  modulated by an appropriate function of the inflaton, as in eq \eqref{Snoinfl},   allows us 
         to obtain the preferred homogeneous profile for $\varphi$ to match the conditions  to impose for a consistent 
             magnetogenesis
          scenario. Having
           a  concrete  example 
           at hands, like this one,  can allow one  to 
            ask whether our ideas pass more refined 
            requirements to have
            consistent magnetogenesis, as 
            for example
             the full dynamics of inflationary fluctuations. 
             This is
             a topic that goes well beyond the scope of this work, but let us emphasize 
              that the dynamics of fluctuations can impose interesting constraints on a given inflationary model, as for example 
              provide an upper bound on the scale of inflation, or can lead to some additional requirements as completing the theory
              in the UV with further operators that control the fluctuation dynamics. This is a broad subject that we intend to pursue in the future.
                        
       \end{appendix}

\end{document}